# The Relationship between the Economic and Financial Crises and Unemployment Rate in the European Union: How Institutions Affected Their Linkage?

**Ionuț JIANU**

Bucharest University of Economic Studies, Bucharest, Romania

ionutjianu91@yahoo.com

Received date:9 March 2019; Accepted date:29 April 2019; Published date: 22 May 2019.

Academic Editor: Andreea Claudia Șerban

**Abstract**

This paper aims to estimate the impact of economic and financial crises on the unemployment rate in the European Union, taking also into consideration the institutional specificities, since unemployment was the main channel through which the economic and financial crisis influenced the social developments.. In this context, I performed two institutional clusters depending on their inclusive or extractive institutional features and, in each cases, I computed the crisis effect on unemployment rate over the 2003-2017 period. Both models were estimated by using Panel Estimated Generalized Least Squares method, and are weighted by Period SUR option in order to remove, in advance the possible inconveniences of the models. The institutions proved to be a relevant criterion that drives the impact of economic and financial crises on the unemployment rate, highlighting that countries with inclusive institutions are less vulnerable to economic shocks and are more resilient than countries with extractive institutions. The quality of institutions was also found to have a significant effect on the response of unemployment rate to the dynamic of its drivers.

---

**Introduction**

The economic and financial crises, which their effects began to spread in the European Union since the second quarter of 2008, have generated new social challenges for the citizens. Unemployment was one of the main channels by which the economic and financial crises have affected the income of the population. However, the effects of the 2009 financial shock were extremely heterogeneous in the European Union member states, the impact of the crisis on unemployment ranging from slight increases followed by steady declines (the case of Germany) to exponential growth rates, was recorded in Greece and Spain as a consequence of high increases in youth unemployment. Moreover, the shock length have been different among member states, in some countries the shocks were absorbed easiest than in others using ineffective absorption mechanisms.

The motivation for choosing this theme consists in the numerous debates focusing on the effects of the economic and financial crises on unemployment, and in the fact that no comparative impact analysis using institutional criterion has been carried out so far.

The main objective of the paper is to provide the specificities regarding the reactivity of unemployment to the shock of the economic and financial crises from the two institutional models of the European Union (the extractive model and the inclusive one). In this context, I proposed to analyse the impact differences between the clusters defined above and the underlying causes. The analysis also took into account other determinants of unemployment, such as youth unemployment, economic growth, and percentage change in nominal unit labor cost.

**Literature Review**

Acemoglu and Robinson (2012) have shown that poverty and inequality are largely determined by the quality of institutions and have separated this concept into extractive and inclusive institutions. In their view, the inclusive institutions represent the written or unwritten rules, by which resources are used for inclusive purposes and create new exploitation opportunities for the population. On the other hand, extractive institutions are seen as those rules that facilitate the extraction of wealth from the population and transfering it into the hands of a limited number of people.

Nicolescu (2016) assessed the security and terrorism related challenges in the EU member states and identified Germany and France as countries that have been confronted with terrorism issues both before and after the refugee crisis . The costs incurred by the private sector as a result of terrorism acts is a relevant sub-indicator of the institutions pillar published by World Economic Forum.

Eurofund (2017) analysed pre / post-crisis social developments and found that unemployment is the main channel by which the financial crisis has affected the social situation in the European Union member states. In this respect, Jianu (2017) identified a positive effect of the unemployment rate on the evolution of the Gini coefficient (indicator used for measuring the income inequality level) in the EU-15 over the 1995-2014 period.

Duca et al. (2017) developed a database of systemic crises and high financial stress periods, covering all the countries from the European Union and Norway for the 1970-2016 period, their methodology was endorsed by a significat number of experts on financial stability. Authors' research separates the acute period of the crisis (ranging from the beginning of the crisis to the last measures adopted to address the effects of the shock) from the post-crisis adjustment period, ending with the normality period. According to them, a systemic crisis represents a shock satisfying at least three conditions as follows:

- ✓ the financial system amplifies the emerged financial shocks;

- ✓ systemically important financial intermediaries face the default risk;
- ✓ responsible bodies intervene strongly by adopting policies for limiting the effects of crises.

The periods of high financial stress (residual events) were included in the database when the shock does not meet the conditions mentioned above, but the financial stress index indicates new financial risks for the evaluated country. High financial stress periods were included in the crises database if economic contractions were recorded for at least 6 consecutive months and high levels of financial stress were identified for one year.

Choudhry et al. (2012) assessed the impact of the financial crisis on youth unemployment over the 1980-2005 period using the Arellano-Bond method in a dynamic panel for 70 countries and identified a higher impact of financial crises on youth unemployment than the one exerted by economic growth on it. The authors represented the financial crises dummy variable according to their classification in the following categories: bank crisis, non-systemic banking crisis, currency crisis and debt crisis.

Calvo et al. (2012) have investigated the implications of the financial crises on the labor market situation and found that those were followed by unemployment recoveries in the case of low inflation. The research is in line with the conclusions of Boeri et al. (2013) stating that the financial crises increase labour market volatility. In a later study, Boeri and Jimeno (2015) highlited that asymmetric shocks had extremely heterogeneous effects on the labour markets in the European Union member states, depending on cross-country institutional differences.

Ştefan (2014) analysed the labour market developments within the European Social Model and found that countries providing high inactive population support are also amongst the countries with the highest employment rates, the low employment rates recorded in other countries are not related to the social benefits magnitude. Postoiu and Avram (2016) investigated the phenomenon of territorial development at European Union level using a set of five relevant indicators, including the employment rate and identified an increase in regional disparities regarding this indicator in recent years.

On the other hand, Ştefan (2011) analysed the developments of unemployment in Romania during 2007-2011 and concluded that youth was the most affected category by unemployment during this period, this is also one of the most vulnerable categories in the presence of economic shocks.

The most important contribution to the investigation of the relationship between unemployment and GDP is held by Okun (1970), which identified an inverse relationship between these variables. Subsequently, a series of studies investigated and accepted the validation of Okun's law in the US or other OECD countries (Prachowny, 1993; Erber, 1994; Padalino and Vivarelli, 1997; Baker and Schmitt, 1999; Lee, 2000 etc). On the other hand, Solow (2000) has shown that unemployment recorded in the European Union member states is a consequence of low demand.

**Methodology**

The methodology used aims to facilitate the estimation of the impact of financial shocks on the unemployment rate in the European Union. In the first phase, the start was from Acemoglu and Robinson's hypothesis according to which poverty and inequality is determined by the quality of the institutions. They divided the institutions into two categories, depending on the performance of the countries in this field: countries with inclusive institutions and countries with extractive institutions.

Firstly, based on the hypothesis that unemployment was one of the main channels by which the economic crisis has affected the people at risk of poverty rate and has led to increased income inequalities, the EU clustering in member states with inclusive and extractive institutions was performed. Subsequently, the impact of financial shocks on the unemployment rate for both resulted groups was estimated, this approach has the cappacity to bring an added value into this research area as it facilitates the comparision between institutional models in terms of unemployment reactivity.

In order to perform the institutional clustering of the European Union, the main institutions indices for all EU member states published by the World Economic Forum in the Global Competitiveness Report 2017-2018 were used, mentioning that the indices published in this report used the institutions score for 2016.

The institutions index published by the World Economic Forum is calculated on the basis of a survey aimed at analysing several qualitative aspects of institutions on a scale from 1 (minimum level) to 7 (maximum), with the exception of the strength of investors protection sub-indicator which was extracted from the World Bank. For a better view into details, the the institutional pillar sub-indicators have been attached in *Table 1*.

In this context, the median of the institutions scores identified for the EU member states was computed in order to facilitate the clustering of the European Union in groups of countries with inclusive and extractive institutions. Thus, the countries that recorded an indicator below the median of 4.29, were categorized as member states with extractive institutions, and those with higher institutions scores than the median were placed in the category of member states with inclusive institutions.

**Table 1: Structure of the institutions pillar**

| Institutions pillar | | | |
|---|---|---|---|
| 1. Property rights | 7. Favoritism in decisions of government officials | 13. Business costs of terrorism | 19. Efficacy of corporate boards |
| 2. Intellectual property protection | 8. Efficiency of government spending | 14. Business costs of crime and violence | |
| 3. Diversion of public funds | 9. Burden of government regulation | 15. Organized crime | 20. Protection of minority shareholders interests |
| 4. Public trust in politicians | 10. Efficiency of legal framework in setting disputes | 16. Reliability of police services | |
| 5. Irregular payments and bribes | 11. Efficiency of legal framework in challenging regulations | 17. Ethical behaviour of firms | 21. Strength of investor protection |
| 6. Judicial independence | 12. Transparency of government policymaking | 18. Strength of auditing and reporting standards | |

*Source: Own processings using World Economic Forum data, The Global Competitiveness Report 2017-2018*

Secondly, statistical data with annual frequency was extracted from Eurostat in order to facilitate the estimation of the impact of financial shocks on the unemployment rate in the European Union. 2003-2017 was used as an analysis period for surprising three distinct economic periods as follows:

1) pre-crisis period (2003-2007);

2) three crisis relevant periods: the overheating, the peak of the economic and financial crisis, the persistence of the crisis effects (2008-2012);
3) post-crisis period (in most countries, reducing the intensity of the crisis effects and the overcoming the challenges posed by the crisis, have occured in 2013).

Financial shocks were captured by a dummy variable constructed on the basis of a review of the European Central Bank assessing the periods of systemic crisis and high financial stress. In this research, the authors separated the residual events and systemic crisis into events related to:

- currency / balance of payments / capital inflows crises - crises due to currency shocks and balance of payments leading to foreign capital outflow, which reduces the money supply.
- crises on the materialisation of sovereign risks - the country risk that highlights the probability of international financial losses caused by macroeconomic or political events.
- macroprudential relevant crises that correspond to high financial stress, but do not meet the criteria for identifying a systemic crisis.
- banking crises caused by high credit and liquidity risks in the banking sector.
- significant asset price correction crises resulting from significant changes in asset prices on bond, currency and real estate markets.
- crises related to transition from a centrally planned economy to a market based economy. However, these events correspond to the developments from the Central and Eastern European countries of the 1990s.

The reason why it has been taken into consideration all types of crisis as well as the high financial stress episodes mentioned in the methodology consists is the fact that these events have the capacity to control the level of unemployment through the propagation of the economic pessimism or by harming other macroeconomic indicators that are in a reverse relationship with unemployment. As a cut-off time for the crisis, the data from the ECB study was used with respect to the deadline for experiencing acute effects of the crisis.

Following the extraction of the unemployment rate and the processing of the dummy variable, a choice was made to control the dependent variable by using other relevant drivers in order to increase the robustness of the estimate.

Thirdly, the stationarity of the data was checked and the most appropriate estimation method was selected. Given that the variables used have been shown to be stationary at both the initial and the first level of integration, a variable highly correlated with the autoregressive term (youth unemployment rate lagged by one year ) was included in the model.

The impact was estimated by applying Panel Estimated Generalized Least Squared (EGLS) method, weighted with Period SUR option for ex-ante removal of heteroskedasticity and general correlations between cross-sections. Also, the Period SUR option increases the value of the Durbin Watson test, which facilitates the validation of the hypothesis of no autocorrelation between residuals..

The specified method was used to estimate the coefficients of the following equation:

$$unem = \alpha_0 + \alpha_1 youth(-1) + \alpha_2 growth + \alpha_3 nomulc + \alpha_4 dummy + \varepsilon_t \quad (1)$$

, where:

- ✓ *unem* = annual rate of unemployment;
- ✓ *youth(-1)* = annual youth unemployment rate, lagged by one;
- ✓ *growth* = annual economic growth rate;
- ✓ *nomulc* = percentage change of nominal unit labour cost (per person);

- ✓ *dummy* = 0 if the economy is in a normal period and financial shocks are absent; *dummy* = 1 if the economy is experiencing financial shocks (systemic crisis or residual events)

Following the estimation of the equation (1), a number of 196 observations from 210 possible observations (for N = 14 and T = 15) has been resulted. The same method for both EU institutional clusters was used, the ultimate objective is to analyse the specifities of each model, either inclusive or extractive. Finally, the characteristics of the residuals, as well as the models adequacy have been checked in order to verify the accuracy of the estimators.

**Results and Interpretations**

This section describes the main results of the model and highlights the different effects of financial shocks on the unemployment rate across the European Union, depending on the member states specificities. Following the analysis of the Institutions main indicator, the clusters mentioned in the methodology were build: (i) the cluster of member states with inclusive institutions; (ii) cluster of member states with extractive institutions (*Table 2*). The results of clustering are largely in line with the structure of the European Union's submodels.

The extractive institutions cluster is composed by member states of catching-up and Southern submodels with some exceptions. Only one country from the catching-up submodel is a part of inclusive institutions cluster (Estonia) and Malta and Portugal, which belong to the Southern submodel of development, and occupy the last two positions in that group of countries. In this context, there is a strong correlation between the level of development of the economies and institutional conditions. On the other hand, besides the aforementioned exception, the cluster of member states with inclusive institutions is composed by countries that are part of the Continental, Northern and Anglo-Saxon submodels.

**Table 2: Institutional clustering of the European Union**

| *Member states with inclusive institutions (score > EU institutions median: 4.29)* | | | | | |
|---|---|---|---|---|---|
| **Country** | **Indicator** | **World Rank / 137** | **EU Rank / 28** | **Score** | **Scale** |
| **Finlanda** | Institutions | 1 | 1 | 6.16 | (1-7) |
| **The Netherlands** | Institutions | 7 | 2 | 5.76 | (1-7) |
| **Luxembourg** | Institutions | 8 | 3 | 5.74 | (1-7) |
| **Sweden** | Institutions | 11 | 4 | 5.59 | (1-7) |
| **United Kingdom** | Institutions | 12 | 5 | 5.52 | (1-7) |
| **Denmark** | Institutions | 13 | 6 | 5.46 | (1-7) |
| **Ireland** | Institutions | 19 | 7 | 5.35 | (1-7) |
| **Germany** | Institutions | 21 | 8 | 5.30 | (1-7) |
| **Austria** | Institutions | 22 | 9 | 5.15 | (1-7) |
| **Estonia** | Institutions | 24 | 10 | 5.04 | (1-7) |
| **Belgium** | Institutions | 25 | 11 | 5.02 | (1-7) |
| **France** | Institutions | 31 | 12 | 4.84 | (1-7) |
| **Malta** | Institutions | 38 | 13 | 4.47 | (1-7) |
| **Portugal** | Institutions | 43 | 14 | 4.40 | (1-7) |

| Member states with extractive institutions (score < EU institutions median: 4.29 | | | | | |
|---|---|---|---|---|---|
| **Country** | **Indicator** | **World Rank / 137** | **EU Rank / 28** | **Score** | **Scale** |
| **Cyprus** | Institutions | 51 | 15 | 4.18 | (1-7) |
| **Czech R** | Institutions | 52 | 16 | 4.16 | (1-7) |
| **Lithuania** | Institutions | 53 | 17 | 4.13 | (1-7) |
| **Spain** | Institutions | 54 | 18 | 4.10 | (1-7) |
| **Slovenia** | Institutions | 56 | 19 | 4.05 | (1-7) |
| **Poland** | Institutions | 72 | 20 | 3.84 | (1-7) |
| **Latvia** | Institutions | 82 | 21 | 3.76 | (1-7) |
| **Romania** | Institutions | 86 | 22 | 3.70 | (1-7) |
| **Greece** | Institutions | 87 | 23 | 3.65 | (1-7) |
| **Slovak Republic** | Institutions | 93 | 24 | 3.51 | (1-7) |
| **Italy** | Institutions | 95 | 25 | 3.50 | (1-7) |
| **Bulgaria** | Institutions | 98 | 26 | 3.48 | (1-7) |
| **Hungary** | Institutions | 101 | 27 | 3.46 | (1-7) |
| **Croatia** | Institutions | 102 | 28 | 3.45 | (1-7) |

*Source: Own processings using World Economic Forum data, The Global Competitiveness Report 2017-2018*

After analysing the sub-indicators of the institutional pillar, three categories of situations regarding the clusters composition were found:

1. *The structure of the sub-indicators related clusters is identical to that of the cluster built on the basis of the main indicator of institutions*: property rights, diversion of public funds, public trust in politicians and ethical behaviour of firms.

2. *The structure of the sub-indicators related clusters is similar to that of the cluster built on the basis of the main indicator of institutions, with some exceptions (maximum two replacements in the inclusive institutions cluster):*

   - ✓ intellectual property protection: PT is replaced by CZ;
   - ✓ irregular payments and bribes: MT is replaced by SI;
   - ✓ judicial independence and favoritism in decisions of government officials: MT is replaced by CY;
   - ✓ efficiency of government spending: AT is replaced by CY;
   - ✓ efficiency of legal framework in setting disputes: PT is replaced by LT;
   - ✓ efficiency of legal framework in challenging regulations and transparency of government policymaking: PT is replaced by CY;
   - ✓ reliability of police services: MT is replaced by ES;
   - ✓ strength of auditing and reporting standards: PT and IE are replaced by CZ and SK;
   - ✓ efficacy of corporate boards: PT and MT are replaced by CZ and LT;
   - ✓ protection of minority shareholders interests: PT is replaced by CZ.

3. *The structure of the sub-indicators related clusters is significantly different from that of the cluster built based on the main indicator of institutions, with several member* *states belonging to the inclusive institutions group being relocated to the extractive institutions group (at least 3 relocations):*

- ✓ burden of government regulation: BE, FR, PT;
- ✓ business costs of terrorism: BE, DE, DK, FR, UK;
- ✓ business costs of crime and violence: BE,DE, FR, IE, NL, UK;
- ✓ organized crime: BE, DE, DK, FR;
- ✓ strength of investor protection: BE, DE, FI, LU, NL, PT.

As it can be observed, the structure of the clusters are broadly convergent, except for the last five sub-indicators. The most significant differences between the main and the secondary clusters occurred as a consequence of reverse divergences between member states related to the different treatment of investors and the high business costs of terrorism, crime and violence. This imbalance is raised by the security concerns faced by the main representatives of the European project.

In the European Union, only 4 member states have maintained their position in the main cluster in all cases of the institutional pillar sub-indicators: EL, PL, RO, SE. Moreover, a special attention should be paid to FI and EE that were in the cluster of member states with extractive institutions for only one sub-indicator of the 21 institutional pillar sub-indicators. In these countries, eGovernment plays an important role in the public administration system, which streamlines the governance process and reduces negative externalities on society.

The efficiency of governance is not determined by the size of the government sector measured by the number of civil servants in the central public administration and is rather a consequence of the political factor, technological progress, cultural factors, the orientation towards e-governance and transparency. All these factors ultimately affect the institutional position of a country, starting from the institutional pillar sub-indicators. The share of civil servants in the central government in the total employed population is quite different and cannot be explained by the developement submodel or by the institutional cluster to which the economy belongs. For example, according to Eurostat, in 2017, MT had the highest level of the indicator (11.8%), while at the opposite end were DE and FR with 0.04% and 0.10%. DK (0.38%), FI (1.13%), NL (1.20%) and SE (1.27%) that are considered to be the most effective governance systems according to the World Bank , have a balanced share of the number of civil servants in the total population, compared to the rest of the EU member states, close to the median of the indicator, except the DK.

Regarding the unemployment rate, in 2017, the highest levels were recorded in EL, ES, IT, CY and HR, which reported unemployment rates in the total active population of over 10% (*Figure 1*). At the end of 2010, the unemployment rate in LT (+12.0 pp), LV (+11.8 pp), EE (+11.2 pp), ES (+ 8.6 pp), SK (+4.9 pp), EL (+4.9 pp), BG (+4.7 pp) and DK (+4.1 pp) was more than 4 percentage points higher than the level recorded in 2008, which also highlights the high impact of the economic and financial crises on this group of countries. However, the impact of the crises has been different in the EU member states. In EL, ES, IT, CY and HR, the impact of the economic and financial crises on unemployment has lasted for a longer period than the crises effects experienced by other member states.

In EL, the unemployment reached its peak in 2013 (27.5%), a moment after which it started to enter an adjustment path, but the 5 year impact length led to an extreme increase in its rate by 19.7 pp compared to the level of the indicator recorded in 2008. The situation was similar in ES and the impact of the economic shock also lasted for 5 years, but the increase in unemployment compared to the level recorded in 2008, although extremely high (+18.3 pp), was slightly lower than the one experienced by EL. However, the impact of high unemployment on GDP was higher in EL than the one manifested in ES, a situation also caused by the higher share of the employed population with temporary contracts in the total employment of the ES. According to Eurostat, in 2009, this indicator stood at 25.3% in ES, while in EL, it was 12.3%. So far, this indicator has increased by 1.5 percentage points in ES and decreased by 0.9 percentage points in EL. On the other hand, in IT, CY and HR, unemployment has begun to enter an adjustment path later, starting with 2015.

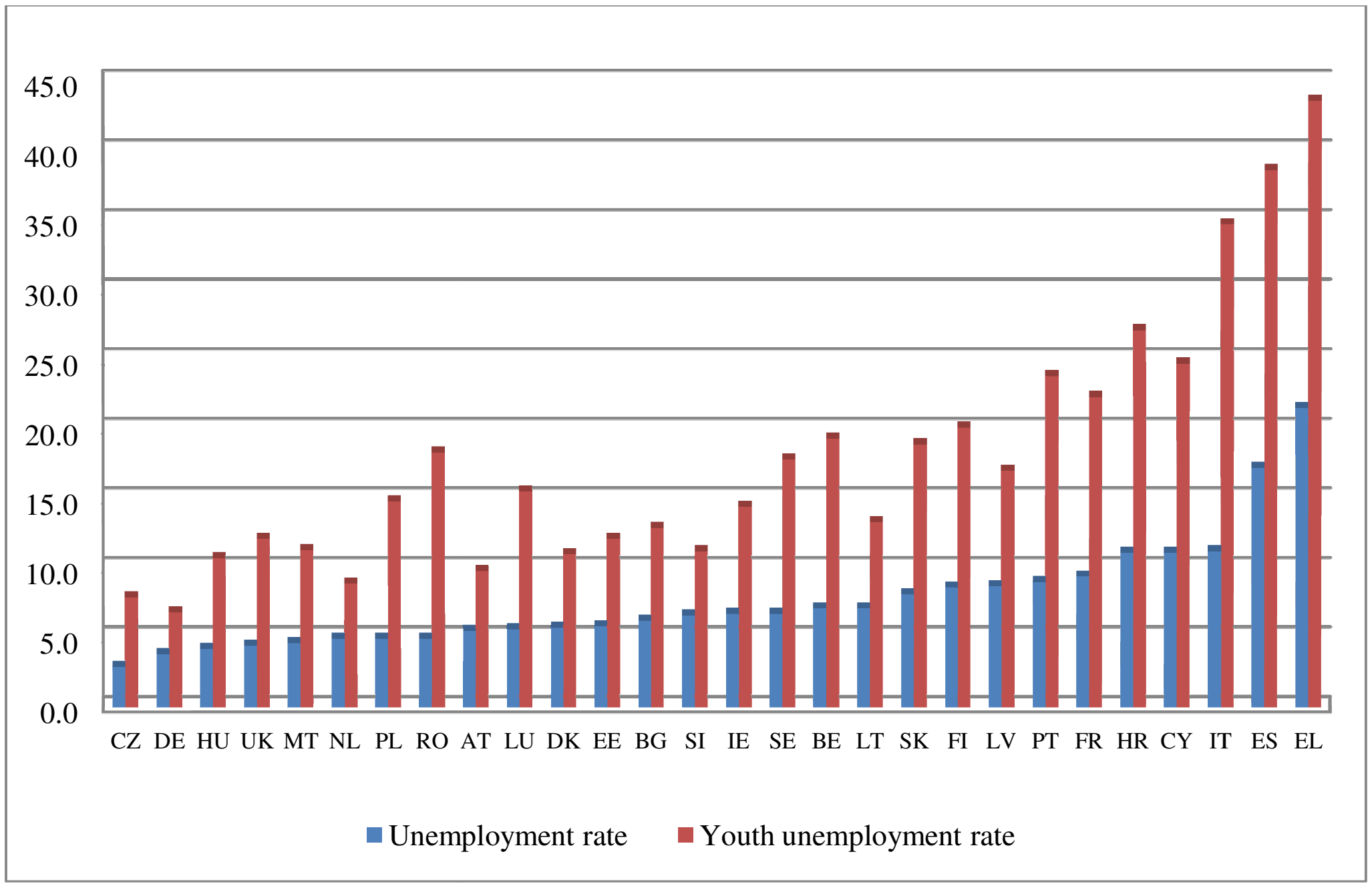

**Figure 1: Unemployment rate in European Union member states in 2017**
*Source: Own processings using Eurostat data*

The shock of the crisis has led to significant restructuring of businesses, which had also unfavorable consequences on the labour market. This group of countries failed to neutralise the shock of the crisis and it was intensified due to the structural problems existing on the labour market, the lack of correlation between educational competences and those needed on the labour market being an important factor in this sense. Although unemployment has been the main channel by which the crisis has brought new social challenges, there have also been happy cases, such as DE, which adequately managed the crisis shock and implemented labour market reforms in order to encourage part-time employment. In DE, the unemployment rate increased in 2009 by only 0.2 percentage points, and it`s on a downward trend starting from 2010.

It can be observed that 8 of the 11 highest unemployment rates in the European Union were recorded in the member states with extractive institutions. Nevertheless, in the extractive institutions cluster, the impact of the crisis on unemployment in RO was the lowest from the EU member states during the 2008-2011 period, increasing by only 1.6 pp, even if the actual conditions shows that CZ, HU and PL are recording lower unemployment rates than RO. In RO, the crisis has affected more the young people participation on labour market, the unemployment of the population aged under 25 increased by 6.4 pp in 2008-2014. In fact, this category was the most vulnerable group at the economic shock of 2009, the lack of jobs for young people still continued to be one of the largest socio-economic problems in RO.

Moreover, the reduced RO unemployment rate is also caused by the high share of persons employed in agriculture, forestry and fishing in the total employment. This can be argued by the fact that RO had the largest share of persons employed in agriculture, forestry and fishing in the total employment in the European Union. In 2017, about a quarter of the employed population (23.7%) was operating in this economic sector. However, the level of the indicator was higher in 2009 (30.1%), which reduced the impact of the economic and financial crisis on unemployment, contrary to a large majority of EU member states. As an indication, the share of employed persons in agriculture, forestry and fishing was 5.4% in 2009 in EU.

A control variable for unemployment is the youth unemployment rate (<25 years), which is also one of the main challenges which European Union is facing in recent years. The most performing member states are DE, CZ and NL with an unemployment rate among young people under 9%, while at the opposite end are IT (34.7%), ES (38.6%) and EL (43.6% %). As can be seen in *Figure 1*, the evolution of the unemployment rate is similar to the one of the youth unemployment rate. The Pearson statistical correlation between these variables was shown to be extremely high and positive during the 2003-2017 period: 92.63% for the cluster of member states with extractive institutions, and 85.53% in the case of the cluster of member states with inclusive institutions. This highlights that there may be found a higher impact of youth unemployment rate on total unemployment in the extractive institutions cluster, given that in this group of countries, young people were most affected by unemployment problems due to incompatibility of skills held with those needed on the labour market.

Regarding the similarity between the evolution of unemployment and that of the economic growth, as well as that of the percentage change in the nominal unit labor costs in the period 2003-2017, negative correlations were found as follows:

- ✓ the statistical correlation between the unemployment rate and the economic growth: the cluster of member states with extractive institutions (-27.84%), the cluster of member states with inclusive institutions (-13.51%);
- ✓ the statistical correlation between the unemployment rate and the percentage change in nominal unit labor costs: the cluster of member states with extractive institutions (-45.07%), the cluster of member states with inclusive institutions (-39.14%).

Although the analysis of the correlation is methodologically important, it should not be confused with the notion of causality, which is more relevant to identify the optimal solutions for addressing indicators imbalances, which is the reason why, following the analysis of the crisis and high financial stress periods, the impact of each variable included in the analysis on unemployment rate was estimated.

Further, the main conclusions of the study conducted by the European Central Bank regarding the periods of manifestation of the effects of economic crises and residual events in the European Union member states have been displayed (*Table 3*).

The economic and financial crises took the most forms in the following EU member states: CY, ES, EL, HR, IE, IT and PT (currency / balance of payments / capital flows crisis, sovereign risk crisis, banking crisis, significant asset price correction crisis and macroprudential crisis). This includes HR and the countries of the southern development model of the European Union, with the exception of MT; a group of member states that faced high challenges as a result of high exposure to the sovereign debt crisis risk. In 2009, the budgetary position worsened in the most of the EU member states excepting MT (which experienced one of the lowest real GDP declines and succeed to significantly reduce the nominal budget deficit) and EE (as a result of the fiscal consolidation program that had a significant contribution on reduction of the share of the budget deficit in GDP higher than the strong decline of real GDP by 14.7%). The most affected countries by this situation began to depend on the loans granted by international creditors, which subsequently led to the sovereign debt crisis. It is worth pointing out that in this country group (CY, ES, EL, HR, IE, IT and PT), the effects still persist even if the acute period of the systemic crisis has ended (except EL).

**Table 3: Crisis and residual events periods in European Union in 2003-2017 period**

| *Country* | *Systemic crisis and residual event periods* | *Country* | *Systemic crisis and residual event periods* |
|---|---|---|---|
| **AT** | 2007-2016 | **IE** | 2008-2013 |
| **BE** | 2007-2012 | **IT** | 2008-2011<br>2011-2013 |
| **BG** | 2007-2011 | **LT** | 2008-2009 |
| **CY** | 2011-2016 | **LU** | 2008-2010 |
| **CZ** | 2007-2010 | **LV** | 2008-2010 |
| **DE** | 2003<br>2007-2013 | **MT** | 2009-2012 |
| **DK** | 2008-2013 | **NL** | 2003-2004<br>2008-2013 |
| **EE** | 2009-2010 | **PL** | 2007-2009 |
| **ES** | 2009-2013 | **PT** | 2008-2015 |
| **FI** | 2008-2010 | **RO** | 2007-2010 |
| **FR** | 2003<br>2008-2009<br>2011-2013 | **SE** | 2008-2010 |
| **EL** | 2010 - ongoing | **SI** | 2009-2014 |
| **HR** | 2007-2012 | **SK** | 2009-2010 |
| **HU** | 2008-2010 | **UK** | 2007-2010 |

*Source: Own processings using European Central Bank Occasional Paper Series, A new database for financial crises in European countries, 2017.*

In order to identify the appropriate method for estimating the impact of financial shocks on the unemployment rate, the Redundant Fixed Effects and the Hausman tests were used to check the compatibility with the fixed or random effects method. The result of the Redundant Fixed Effects test (Likelihood Ratio) led to the rejection of the null hypothesis consisting in the fact that this method is redundant and to the acceptance of the alternative one that supports the fixed-effect model as an estimation method (*Table 4*).

**Table 4: Tests performed**

| *Test* | *Assumption checked* | *Probability** | *Result* |
|---|---|---|---|
| Redundant Fixed Effects - Likelihood Ratio test | Fixed effects model is redundant | I - 0.00% | Fixed effects model is indicated |
| | | E - 0.00% | Fixed effects model is indicated |
| Correlated Random Effects - Hausman test | Compatibility with random effects model | I - 40.45% | Random effects model is indicated |
| | | E -67.17% | Random effects model is indicated |
| Histogram - Normality test (Jarque-Bera)** | Normal distribution of the residuals | I - 11.29% | Residuals are normally distributed |
| | | E - 07.76% | Residuals are normally distributed |
| Cross-section dependence test | Absence of cross-section dependence | I - 100,00% | No cross-section dependence |
| | | E -100,00% | No cross-section dependence |
| (Breusch - Pagan) | (1) | | |
| Cross-section dependence test (Pesaran CD) | Absence of cross-section dependence (2) | I - 83,16% | No cross-section dependence |
| | | E -53.37% | No cross-section dependence |
| Breusch - Pagan - Godfrey Heteroskedasticity test | Homoskedasticity | I - 58.66% (I*) | Model is homoskedastic |
| | | E - 31.82% (E*) | Model is homoskedastic |
| Breusch - Pagan serial correlation test | Absence of serial correlation | I - 18.00% (I**) | No serial correlation |
| | | E - 08.21% (E**) | No serial correlation |

* I = inclusive institutions cluster; E = extractive institutions cluster.
** see *figure 2* for a detailed picture.
I* for n = 196 and degrees of freedom = 4; E* for n = 195 and degrees of freedom = 4.
I** for n = 168 and degrees of freedom = 2; E** for n = 167 and dregrees of freedom = 2.

*Source: Own calculations using Eviews 9.0*

On the other hand, the result of the Hausman test that checks the compatibility with the random effects method indicated the acceptance of this method as the most appropriate estimation technique. The indecision resulting from the two tests led to the use of a standard model without fixed or random effects. This also enabled the application of the Period SUR option, which operates the necessary adjustments to ex-ante removal of cross-section dependence and heteroskedasticity, as well as correcting the Durbin-Watson value, thereby facilitating the confirmation of the hyphothesis of no autocorrelation between residuals.

**Inclusive institutions cluster**

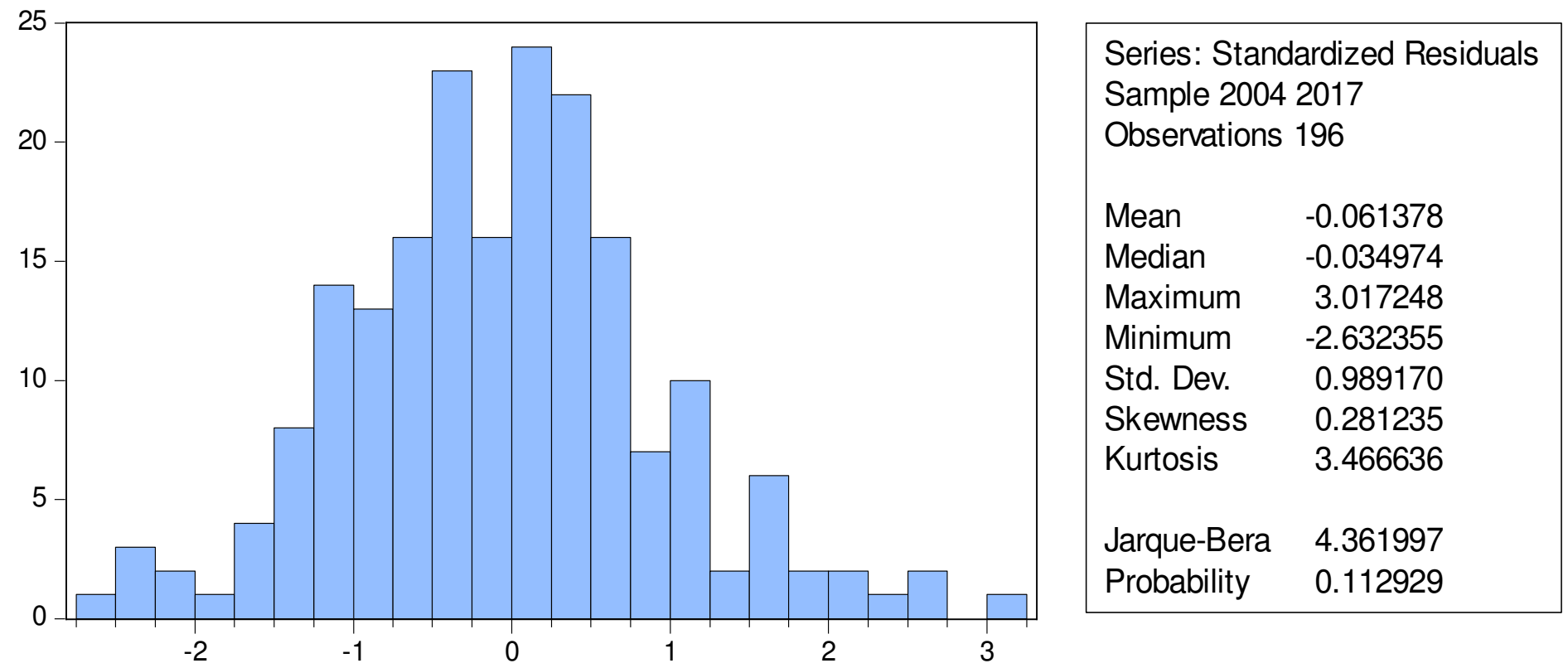

**Extractive institutions cluster**

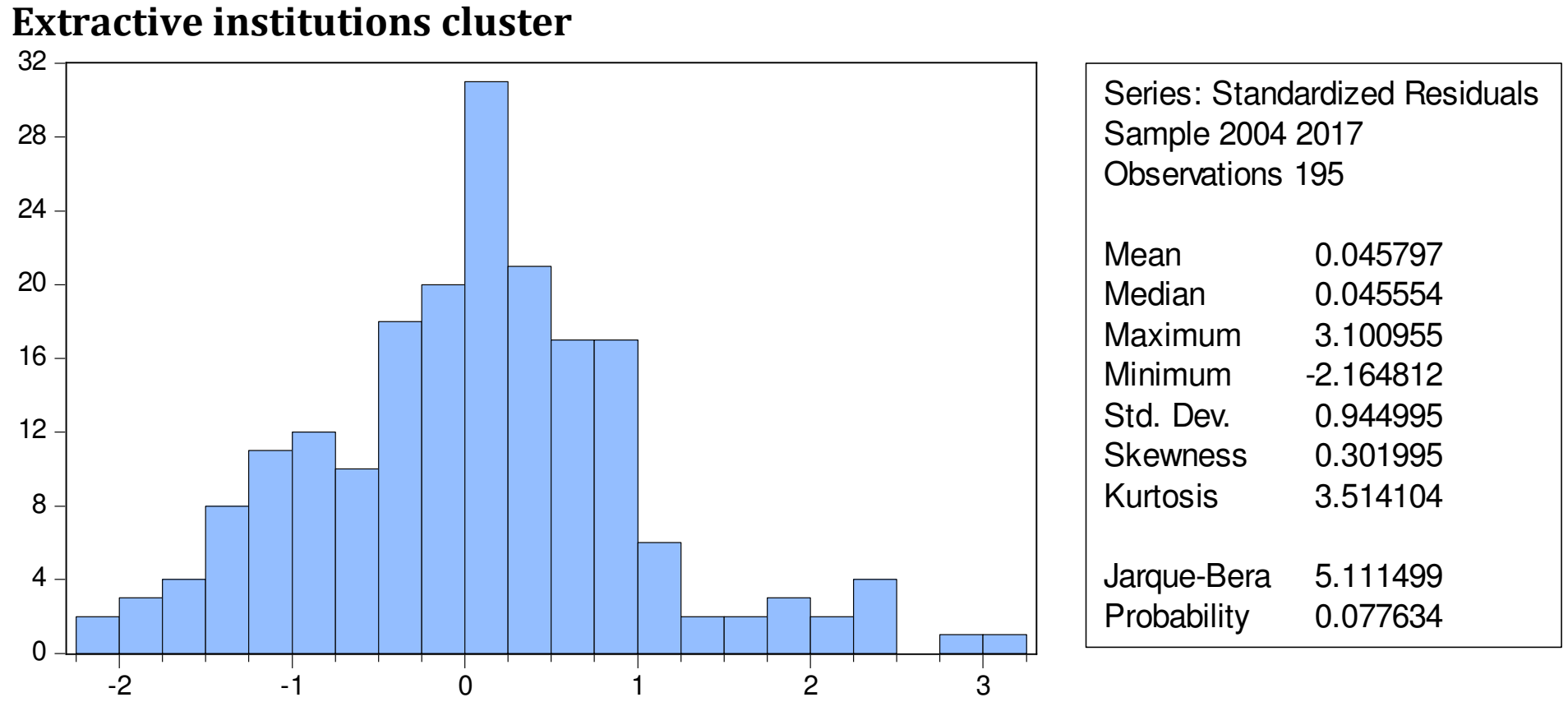

**Figure 2: Histogram - Normality test**

*Source: Own calculations using Eviews 9.0*

According to *Figure 3,* the probability associated to Fisher test demonstrates the statistical validity of both models, while the differences between the degrees of which unemployment is explained by the variables used in the analysis are minor. The unemployment rate corresponding to the cluster of member states with inclusive institutions is explained in a higher proportion by the determinants used (98.98%) than the situation indicated by the R-squared value computed for the model of extractive institutions cluster (93.53%). Moreover, all the coefficients have proven to be statistically significant, which creates the premises of accepting their verosimility.

The 1 pp increase in the youth unemployment rate recorded in the previous year leads to an increase in the unemployment rate from the current year by 0.38 percentage points in the cluster of member states with extractive institutions, which is higher than the impact estimated for the extractive institutions cluster. In principle, a large proportion of young unemployed face high labour market challenges, they risking to have the same status on labour market after one year, which makes them a vulnerable social category, likely to remain in unemployment for a long time. For example, the highest long-term unemployment rates among young people in 2017 were recorded in EL (53.9%), IT (53.8%), SK (45.2%) and RO (38, 0%), while FI (5.8%), SE (5.8%), DK (6.4%) and NL (13.9%) recorded the lowest level of this indicator from the EU.

Model. Inclusive institutions cluster
Dependent Variable: UNEM
Method: Panel EGLS (Period SUR)
Sample (adjusted): 2004 2017
Periods included: 14
Cross-sections included: 14
Total panel (balanced) observations: 196
Linear estimation after one-step weighting matrix
Period SUR (PCSE) standard errors & covariance (d.f. corrected)

| Variable | Coefficient | Std. Error | t-Statistic | Prob. |
|---|---|---|---|---|
| YOUTH(-1) | 0.290796 | 0.004714 | 61.68981 | 0.0000 |
| GROWTH | -0.218151 | 0.002434 | -89.63627 | 0.0000 |
| NOMULC | -0.252014 | 0.004197 | -60.04302 | 0.0000 |
| DUMMY | 0.422175 | 0.028301 | 14.91726 | 0.0000 |
| C | 3.133585 | 0.113609 | 27.58207 | 0.0000 |

Weighted Statistics

| | | | |
|---|---|---|---|
| R-squared | 0.989883 | Mean dependent var | 1.969115 |
| Adjusted R-squared | 0.989671 | S.D. dependent var | 13.62849 |
| S.E. of regression | 1.001406 | Sum squared resid | 191.5375 |
| F-statistic | 4672.099 | Durbin-Watson stat | 1.935432 |
| Prob(F-statistic) | 0.000000 | | |

Unweighted Statistics

| | | | |
|---|---|---|---|
| R-squared | 0.746510 | Mean dependent var | 7.412245 |
| Sum squared resid | 331.2072 | Durbin-Watson stat | 0.381893 |

Model. Extractive institutions cluster
Dependent Variable: UNEM
Method: Panel EGLS (Period SUR)
Sample (adjusted): 2004 2017
Periods included: 14
Cross-sections included: 14
Total panel (unbalanced) observations: 195
Linear estimation after one-step weighting matrix
Period SUR (PCSE) standard errors & covariance (d.f. corrected)

| Variable | Coefficient | Std. Error | t-Statistic | Prob. |
|---|---|---|---|---|
| YOUTH(-1) | 0.383801 | 0.007791 | 49.25973 | 0.0000 |
| GROWTH | -0.174762 | 0.014213 | -12.29627 | 0.0000 |
| NOMULC | -0.052466 | 0.007394 | -7.095362 | 0.0000 |
| DUMMY | 0.820551 | 0.115692 | 7.092538 | 0.0000 |
| C | 1.052700 | 0.252978 | 4.161224 | 0.0000 |

Weighted Statistics

| | | | |
|---|---|---|---|
| R-squared | 0.935369 | Mean dependent var | 1.269520 |
| Adjusted R-squared | 0.934008 | S.D. dependent var | 4.765687 |
| S.E. of regression | 0.956017 | Sum squared resid | 173.6539 |
| F-statistic | 687.4397 | Durbin-Watson stat | 1.937385 |
| Prob(F-statistic) | 0.000000 | | |

Unweighted Statistics

| | | | |
|---|---|---|---|
| R-squared | 0.816078 | Mean dependent var | 10.57231 |
| Sum squared resid | 892.4774 | Durbin-Watson stat | 0.324688 |

**Figure 3:** ***Results of the models***

*Source: Own calculations using Eviews 9.0*

The impact differences between the models are caused by the problems of non-correlation between educational skills and necessary competences on the labour market, which are felt with a higher intensity in the cluster of member states with extractive institutions. Moreover, there are cultural differences between the clusters reviewed, the youngsters belonging to the extractive institutions cluster of states are more dependent on parents, while the young people from the developed countries are being stimulated by families to join on the labour market and to become independent from a younger age, which means that the length of staying in unemployment category is lower for this group. Also, higher salary expectations than those provided by the economic agents operating in the cluster of member states with extractive institutions can be brought into discussion. This pattern of development favors more significant and short-term cash gains than the population of member states with inclusive institutions.

As regards the relationship between economic growth and unemployment, the reverse relationship between these variables was confirmed in both cases, the impact is higher for the group of member states with inclusive institutions (0.21 pp) than the one calculated for the model of member states with extractive institutions (0.17 pp). This highlights the fact that the relationship between economic growth and unemployment rate is more robust in the model analysing extractive institutions. One of the main factors contributing to this situation is the low flexibility of the labour market in this category of states, which limits the reduction of unemployment even in the conditions of high rates of economic growth. As a result of the labour market flexibility weaknesses, the countries with extractive institutions fail to adequately stimulate the transfer of the human labor factor from regions with low unemployment rates to those facing problems of sub-production and unemployment. This situation is reflected also in the high regional disparities in this group of countries compared to those experienced by the member states with inclusive institutions.

According to the results shown in *Figure* 3, in the period 2003-2017, the increase in the percentage change in nominal unit labor costs per person employed by 1 pp led to a decrease in the unemployment rate by 0.25 percentage points in the member states with inclusive institutions and only by 0.05 in the group of countries with extractive institutions. The negative relationship is explained by the important role of the wage level and other benefits offered to the employed population in stimulating people's participation on the labour market. The lower impact of the cluster of extractive institutions can be explained by the fact that the 1 percentage change in a higher salary is higher in nominal terms than the 1 percentage change in a lower salary, and this does not make the labour market attractive for the population living in member states with extractive institutions.

From the point of view of the impact of financial shocks, the member states with extractive institutions were less resilient. In this group, the effect of an economic shock experienced in one year led to an increase in unemployment by 0.82 pp in 2003-2017 period, almost double than the impact found in the case of countries with inclusive institutions (0.42 pp). Financial shocks have the potential to increase unemployment through multiple channels: job cuts due to a consequence of spreading the economic pessimism, lower investments and economic contraction, as well as benefit and wage cuts. Stimulating part-time employment as well as using the concept of flexicurity in the Northern countries as well as in DE have improved the resilience of the labour market to the effects of the economic and financial crises. The lack of structural reforms has also caused the higher reactivity of unemployment in the case of the cluster of member states with extractive institutions.

The standardized residual series of the two estimated models have proven to be normally distributed, given that the Jarque-Bera test probability exceeded the 5% significance threshold in both cases. Both tests run to verify the cross-section dependence between the residuals have provided probabilities of over 5%, which indicates the acception of the assumption of no cross-section dependence in both models. Also, the tests run to check heteroskedasticity and autocorrelation of the residuals confirmed the hypothesis of homoskedasticity and no serial correlation. Regarding multicolinearity, it was confirmed its absence as R-squared value is higher than the correlation coefficients recorded between independent variables.

Taking into account the results of the tests performed, the appropriate representation of the models was accepted, as well as their coefficients and their sign.

## Conclusions

The clustering of the European Union member states depending on the institutional pillar has been largely in line with the development phase and with the intensity and the duration of the financial shocks in the European economies. Essentially, the inclusive institutions cluster is composed of the Nordic, Continental and Anglo-Saxon countries, only Estonia (the catching-up submodel) and Portugal (the Southern submodel) joining this group, these are only a part of other submodels of the EU. Except for the two countries mentioned above, the extractive institution cluster is composed strictly of the member states of the Southern and the catching-up submodels.

The impact of the financial shocks recorded in the period 2003-2017 on the unemployment rate proved to be much higher in the case of the member states forming the extractive institutions cluster than the impact assessed for the countries with inclusive institutions. This was also caused by the measures implemented actively by the member states with inclusive institutions to improve the economic resilience. In order to reduce the reactivity of unemployment to the emergence of financial shocks, the following are recommended: the improvement of the flexibility of the labour market, the adoption of necessary measures for encouraging a cohesive regional development process, tackling the problems of non-correlation of educational skills with those needed on the labour market, reforms to tackle the insertion of young people on labour market by which this group will be no longer a social category exposed to the long-term unemployment, especially in member states with extractive institutions, which results also from the higher impact of the youth unemployment recorded in the previous year on the current unemployment which is found in the case of extractive institution model.

It is also recommended to pay a higher attention to the sub-indicators of the institutional pillar published by World Economic Forum and to approach these problems in order to moderate the unemployment. It is worth mentioning that this phenomenon can be misleading and other labour market specificities as the employment structure, respectively the size of the population employed in agriculture, forestry and fishing or the population working in sectors without high value added should be taken into account.